\documentclass[twocolumn,showpacs,preprintnumbers,amsmath,amssymb]{revtex4}

\usepackage{graphicx}
\usepackage{dcolumn}
\usepackage{bm}

\begin{document}

\title{Effective Action and Schwinger Pair Production in Scalar QED}
\author{Sang Pyo Kim}\email{sangkim@kunsan.ac.kr}
\affiliation{Department of Physics, Kunsan National University,
Kunsan 573-701, Korea} \affiliation{Asia
Pacific Center for Theoretical Physics, Pohang 790-784, Korea}

\author{Hyun Kyu Lee}\email{hyunkyu@hanyang.ac.kr}
\affiliation{Department of Physics and BK21 Division of Advanced
Research and Education in Physics, Hanyang University, Seoul
133-792}

\date{}

\begin{abstract}
Some astrophysical objects are supposed to have very strong
electromagnetic fields above the critical strength. Quantum
fluctuations due to strong electromagnetic fields modify the Maxwell
theory and particularly electric fields make the vacuum unstable
against pair production of charged particles. We study the strong
field effect such as the effective action and the
Schwinger pair production in scalar QED. 
\end{abstract}

 \maketitle

\section{Introduction}

Gravitational interaction and electromagnetic interaction are the
most important fundamental interactions in astrophysics. The
non-relativistic Newtonian gravity is no longer a proper theory for
strong gravity. The Einstein gravity, a new paradigm, replaces the
Newtonian gravity and predicts entirely new phenomena such as black
holes when the gravitational interaction is extremely strong. One
may wonder how the Maxwell theory would be modified for extremely
strong sources of charges and current or external electromagnetic
fields. As the Maxwell theory is already a relativistic theory for
electromagnetic phenomena, it is not likely that a new theory would
replace the Maxwell theory even for reasonably strong sources or
electromagnetic fields up to the unification scale for gravity and
electromagnetism. Thus, below the unification scale, the Maxwell
theory is still the building block for quantum phenomena.

Virtual pairs of charged particles experience the interaction with
the background of strong electromagnetic fields and undergo quantum
fluctuations contributing to the effective action. One remarkable
phenomenon due to a strong electromagnetic field, in particular,
strong electric field, is Schwinger pair production of charged
particles from the vacuum \cite{Schwinger}. In fact, the virtual
particle and antiparticle from the Dirac sea can gain a sufficient
kinetic energy comparable or greater than the rest mass energy over
the Compton wavelength and then tunnel the potential barrier from
the Dirac sea to substantiate into real pairs. The critical strength
for electron-positron pair production is $E_c = m_e^2 c^3/e
\hbar~(1.3 \times 10^{16} {\rm V/cm})$.

A particle with charge $q$ and mass $m$ in a constant magnetic field
undergoes circular motions with the Larmor frequency $\omega_c =
qB/mc$. In quantum theory charged particles occupy the Landau levels
with energy $E = \hbar \omega_c n$ and in a strong magnetic field of
$B_c = m^2 c^2 / q \hbar$ $(4.4 \times 10^{13} {\rm G}$ for an
electron), the energy difference of Landau levels $\delta E = \hbar
\omega_c$ can be comparable to the rest mass of the particle. In the
transverse direction of a magnetic field above the critical strength
electrons of an atom are strongly bounded and fill the lowest Landau
levels but in the parallel direction are attracted by the Coulomb
force. Further, the energy of Landau levels yields a quantum
correction to the Maxwell term. It turns out that the quantum
correction contributes nonlinear terms of Lorentz invariant field
tensors to the effective action and the vacuum thus becomes
polarized effectively similar to nonlinear polarized media
\cite{Heisenberg-Euler,Weisskopf,Schwinger} (For a review on strong
QED, see \cite{Dunne}).

At present the most strong electromagnetic fields are generated not
by terrestrial experiments but by astrophysical sources. Even the
strong electromagnetic fields that would be generated by terrestrial
experiments or International Linear Collider in the future
\cite{Ringwald} are still below the magnetic fields of neutron
stars, which are remnants of massive stars. The magnetic flux of a
normal star with radius $R_0$ and field $B_0$ is frozen during the
collapse and amplified to $B = B_0 (R_0/R_n)^2$. The magnetic field
may be further amplified by dynamos. Neutron stars have magnetic
fields ranging from $10^8 {\rm G}$ to $10^{15} {\rm G}$ \cite{Wood}.
Magnetars have the strongest magnetic field above the critical
strength. The electromagnetic phenomena in magnetars would
significantly differ from the Maxwell theory due to QED effects
\cite{Duncan,Harding-Lai}.

Only with electromagnetic interactions it is hard to accumulate
charges to generate an electric field beyond the critical strength
partly because of the repulsive force among the same kind of charges
and partly because of the decay of the field itself by emitting
pairs. However, strange stars, hypothetical quark stars of quark
matter \cite{Ito,Bodmer,Witten,Farhi}, can hold electrons on the
crust by attractive Coulomb force by positively charged core which
in turn tightly bounded by nuclear force \cite{Alcock,Haensel}. The
electrosphere of strange stars, if exist, can be the strongest
source for electric fields with two order higher than the critical
strength and the emission of charged particles from Schwinger pair
production may be an evidence of strange stars
\cite{Usov98,Usov05,Harko06}.

The purpose of this paper is to study the effective action in scalar
QED in the presence of a strong external electromagnetic field and
the Schwinger pair production due to the vacuum instability. Here we
present a novel method to calculate the effective action
particularly in electric fields. In the Hamiltonian approach
developed in Ref. \cite{Kim-Lee07}, which is based on the invariant
method \cite{Lewis-Riesenfeld}, the evolution operator can be
expressed by a two-mode squeezed operator, in terms of which the
ingoing vacuum evolves to the outgoing vacuum. The scattering
amplitude of the ingoing vacuum to the outgoing vacuum is the
expectation value of the two-mode squeeze operator and the effective
action is thus determined by the Bogoliubov coefficients between the
ingoing and outgoing particle and antiparticle operators. The
adiabaticity of the evolution assumed in Ref. \cite{Ambjorn-Hughes}
is not required for the effective action. Further we find the Fock
space of number states of particles and antiparticles, which evolve
from the ingoing vacuum to the outgoing vacuum.

The organization of this paper is as follows. In Sec. II, we briefly
review the effective action in scalar QED under the influence of
uniform magnetic field and electric field. In Sec. III, we apply the
Hamiltonian approach to study the Fock space for charged particles
and antiparticles under an external electromagnetic field. In Sec.
IV, we find the effective action in terms of the Bogoliubov
coefficients by calculating the scattering amplitude of the ingoing
vacuum to the outgoing vacuum.

\section{Effective Action in Constant B and E}

In this section, following the Hamiltonian approach developed in
Ref. \cite{Kim-Lee07}, we study scalar QED described by the
Klein-Gordon equation for a charged boson with $q~(q
> 0)$ and $m$ (in units with $\hbar = c = 1$ and with metric
signature $(+, -, -, -)$)
\begin{eqnarray}
 [\eta^{\mu \nu} (\partial_{\mu} + i q A_{\mu}) (\partial_{\nu}
   + i q A_{\nu}) + m^2] \phi ({\bf x}, t) = 0. \label{kg fmod}
\end{eqnarray}
The charged boson has the Lagrangian
\begin{equation}
L = \int d^3 x \Bigl[ \eta^{\mu \nu} (\partial_{\mu} - i q A_{\mu})
\phi^* (\partial_{\mu} + i q A_{\mu}) \phi - m^2 \phi^* \phi \Bigr].
\label{lag}
\end{equation}
In a constant magnetic field along the $z$-direction, we choose the
gauge in a symmetric form
\begin{eqnarray}
A_{\mu} = \Bigl( 0, - \frac{By}{2}, \frac{Bx}{2}, 0 \Bigr).
\end{eqnarray}
As classically the charged particle moves along a spiral of circular
motion in the transverse direction and linear motion along the
parallel direction of the magnetic field, we decompose the field
$\phi = \phi_{\perp} (x, y) \phi_z$, with the transverse part being
real and the parallel part being complex. After integrating over the
transverse direction, the corresponding Hamiltonian is given by
\begin{equation}
H = \int dz  \Bigl[\pi^* \pi + (\partial_z \phi_z^*) (\partial_z
\phi_z) + \Bigl( H_{\perp} (x, y) + m^2 \Bigr) \phi_z^* \phi_z
\Bigr], \label{ham-B}
\end{equation}
where $\pi = \dot{\phi}^*$ and $\pi^* = \dot{\phi}$ are the
conjugate momenta, and the transverse part is
\begin{eqnarray}
H_{\perp} (x, y) = p_x^2 + p_y^2 + \Bigl(\frac{qB}{2} \Bigr)^2 (x^2
+ y^2) + (qB) L_z. \label{2d os}
\end{eqnarray}
Here $p_x = - i \partial_x$, $p_y = - i \partial_y$ and $L_z = x p_y
- y p_x$ is the $z$-component of angular momentum. The Klein-Gordon
equation has the solution of the form $\phi ({\bf x}, t) = e^{- i
\omega t + i k_z z} \phi_{\perp} (x, y)$, where
\begin{eqnarray}
\Bigl[ H_{\perp} (x, y) + m^2 + k_z^2 \Bigr] \phi_{\perp} (x, y) =
\omega^2 \phi_{\perp} (x, y),
\end{eqnarray}
Note that the transverse part is a two-dimensional oscillator with
the mass $1/2$ and the frequency $qB$ together with the
$z$-component of angular momentum.

One may use the oscillator representation of $U(3)$, which has
$SO(3)$ as a subgroup, with the generators \cite{wybourne}
\begin{eqnarray}
T_{ij} = \frac{1}{2} (c^{\dagger}_i c_j + c_j c^{\dagger}_i),
\end{eqnarray}
where $c_i$ and $c^{\dagger}_i$ are annihilation and creation
operators for the component $i = x, y, z$. It follows that the
angular momentum takes the form
\begin{eqnarray}
L_z = i (T_{yx} - T_{xy}),
\end{eqnarray}
and the transverse part thus has the oscillator representation
\begin{eqnarray}
H_{\perp} = qB (T_{xx} + T_{yy}) + i (qB) (T_{yx} - T_{xy}).
\end{eqnarray}
In the new basis
\begin{eqnarray}
c_{\pm} = \frac{1}{\sqrt{2}} (c_x \pm i c_y),
\end{eqnarray}
the transverse part takes the diagonal form
\begin{eqnarray}
H_{\perp} &=& qB (T_{++} + T_{--}) + qB (T_{--} - T_{++})
\nonumber\\
&=& 2 (qB) T_{--}.
\end{eqnarray}
The eigenstates of the transverse part are the number states $\vert
n_- \rangle =(c_-^{\dagger})^{n_-} \vert 0_- \rangle/ \sqrt{n_-!}$
with the energy eigenvalue
\begin{eqnarray}
H_{\perp} \vert n_- \rangle  = qB (2n_- +1 ) \vert n_- \rangle.
\label{landau}
\end{eqnarray}

Hence the energy eigenstate and eigenvalue of the Hamiltonian
(\ref{ham-B}) are given by
\begin{eqnarray}
H \vert n, k_z \rangle &=& \omega(n, k_z) \vert n, k_z \rangle
\nonumber\\
&=& \sqrt{m^2 + k_z^2 + qB (2n +1 )} \vert n, k_z \rangle. \label{en
sp}
\end{eqnarray}
The effective action of scalar QED may be found from the scattering
amplitude
\begin{eqnarray}
e^{ i S_{\rm eff}} = \langle {\rm out} \vert {\rm in} \rangle,
\label{sc am}
\end{eqnarray}
where the outgoing state evolves from the ingoing state as
\begin{eqnarray}
\vert {\rm out} \rangle = U \vert {\rm in} \rangle.
\end{eqnarray}
Using the evolution operator for the Hamiltonian (\ref{ham-B}) in
the magnetic field
\begin{eqnarray}
U = e^{- i H t},
\end{eqnarray}
we find the effective action
\begin{eqnarray}
e^{ i S_{\rm eff}} = \prod_{n, k_z} e^{i \omega (n, k_z)} \langle n,
k_z \vert n, k_z \rangle.
\end{eqnarray}
Now the effective action per unit time and unit area, ${\cal L}_{\rm
eff} = S_{\rm eff} /({\rm time}) \times ({\rm area})$,  is given by
\begin{eqnarray}
{\cal L}_{{\rm eff}} = \frac{qB}{(2 \pi)^2} \int_{- \infty}^{\infty}
dk_z \sum_{n = 0}^{\infty} \sqrt{m^2 + k_z^2 + qB (2n +1 )},
\label{ef b}
\end{eqnarray}
where $qB/(2 \pi)$ is the number of Landau levels and $1/(2 \pi)$ is
the number of states for $k_z$. To perform the summation and
integral, following Ref. \cite{greiner94}, we differentiate (\ref{ef
b}) twice with respect to $m^2$ and sum over $n$ to obtain
\begin{eqnarray}
\frac{\delta^2 {\cal L}_{{\rm eff}}}{\delta (m^2)^2} = \frac{qB}{16
\pi^2} \int_0^{\infty} ds e^{- m^2 s} \frac{1}{\sinh (qB s)}.
\end{eqnarray}
Integrating twice with respect to $m^2$, we obtain the effective
action
\begin{eqnarray}
{\cal L}_{{\rm eff}} (B) = - \frac{1}{16 \pi^2} \int_0^{\infty} ds
\frac{e^{- m^2 s}}{s^3} \hspace{0.6 in} \nonumber\\ \times \Bigl[
\frac{qBs}{\sinh (qBs)} - 1 + \frac{(qBs)^2}{6} \Bigr],
\end{eqnarray}
where we fix the integration constants such that ${\cal L}_{\rm eff}
(B = 0) = 0$. The effective action for a constant electric field may
be obtained from the duality $B = iE$
\begin{eqnarray}
{\cal L}_{{\rm eff}} (E) = - \frac{1}{16 \pi^2} \int_0^{\infty} ds
\frac{e^{- m^2 s}}{s^3} \hspace{0.6 in} \nonumber\\ \times \Bigl[
\frac{qEs}{\sin (qEs)} - 1 - \frac{(qEs)^2}{6} \Bigr].
\label{eff-E0}
\end{eqnarray}
In terms of the Lorentz invariant tensors
\begin{eqnarray}
{\cal F} &=& \frac{1}{4} F_{\mu \nu} F^{\mu \nu} = - \frac{1}{2}
({\bf E}^2 - {\bf B}^2)
= \frac{1}{2} (X_r^2 - X_i^2), \nonumber\\
{\cal G} &=& \frac{1}{4} F_{\mu \nu} \tilde{F}^{\mu \nu} = - {\bf E}
\cdot {\bf B} = X_r X_i,
\end{eqnarray}
the effective action can be expressed as
\begin{eqnarray}
{\cal L}_{{\rm eff}} (B, E) = - \frac{1}{16 \pi^2} \int_0^{\infty}
ds
\frac{e^{- m^2 s}}{s^3} \hspace{1.1 in} \nonumber\\
\times \Bigl[ \frac{(q X_r s)(q X_i s)}{\sinh (q X_r s) \sin(q X_i
s)} - 1 - \frac{(q s)^2}{6} (X_r^2 - X_i^2) \Bigr]. \label{ef b2}
\end{eqnarray}
For another derivation of the effective action, see Refs.
\cite{Schwinger,Heisenberg-Euler,Weisskopf}.

The real part of the effective action up to quartic power of fields
is
\begin{eqnarray}
{\rm Re} {\cal L}_{{\rm eff}} (B, E)  = \frac{q^4}{720 \pi^2 m^4} (
4 {\cal F}^2 + 7 {\cal G}^2 ).
\end{eqnarray}
The imaginary part comes from the pole structure of the integral and
takes the form
\begin{eqnarray}
2 {\rm Im} {\cal L}_{{\rm eff}} (B, E) = \frac{1}{8 \pi^3} \sum_{n =
1}^{\infty} (-1)^{n+1} \Bigl( \frac{qX_i}{n} \Bigr)^2 e^{- \frac{\pi
m^2 n}{qX_i}}. \label{sch pair}
\end{eqnarray}
In the pure electric field $X_i = E$ and Eq. (\ref{sch pair}) is the
Schwinger pair production rate for charged bosons.

\section{Fock Space in Scalar QED}

In an electric field in the $z$-direction with the time-dependent
gauge potential $A_{\mu} = (0, 0, 0, A_z (t))$, the Hamiltonian is
given by
\begin{equation}
H(t) = \int d^3 x  \Bigl[\pi^* \pi + \phi^* \Bigl(-
\partial_{x}^2 - \partial_y^2 - (\partial_z + i q A_z)^2 + m^2 \Bigr) \phi
\Bigr]. \label{ham-E}
\end{equation}
The quantum state of the charged particle obeys the functional
Schr\"{o}dinger equation \cite{FHM}
\begin{eqnarray}
i \frac{\partial \Phi (t, {\bf x})}{\partial t}  = H (t) \Phi (t,
{\bf x}).
\end{eqnarray}
In the scattering theory the final state evolves from an initial
state
\begin{eqnarray}
\vert {\rm out} \rangle = U(t_{\rm out}, t_{\rm in}) \vert {\rm in}
\rangle,
\end{eqnarray}
where $U$ is the evolution operator
\begin{eqnarray}
i \frac{\partial U (t)}{\partial t} = H (t) U (t).
\end{eqnarray}

The prominent feature of the Hamiltonian (\ref{ham-E}) is the
explicit time-dependency through $A_z (t)$. For a time-dependent
Hamiltonian, we may use the invariant method, in which invariant
operators satisfying the Liouville-von Neumann equation provide the
exact quantum states \cite{Lewis-Riesenfeld}. Our stratagem is to
find the time-dependent annihilation and creation operators for each
mode of the free scalar field (\ref{ham-E}), which are invariant
operators. As we now have a field instead of a finite system, we
first quantize the field as
\begin{eqnarray}
\phi (t, {\bf x}) &=& \int [d {\bf k}] \Bigl[ \varphi_{\bf k} (t)
a_{\bf k} (t) + \varphi^*_{\bf k} (t) b^{\dagger}_{\bf k} (t) \Bigr]
e^{i {\bf k} \cdot {\bf
x}},\nonumber\\
\phi^* (t, {\bf x}) &=& \int [d {\bf k}] \Bigl[ \varphi_{\bf k} (t)
b_{\bf k} (t) + \varphi^*_{\bf k} (t) a^{\dagger}_{\bf k} (t) \Bigr]
e^{- i {\bf k}. \cdot {\bf x}},
\end{eqnarray}
where $[d {\bf k}] = d^3 k / (2 \pi)^3$, and the momentum as
\begin{eqnarray}
\pi (t, {\bf x}) &=& \int [d {\bf k}] \Bigl[ \dot{\varphi}^*_{\bf k}
(t) a^{\dagger}_{\bf k} (t) + \dot{\varphi}_{\bf k} (t) b_{\bf k}
(t) \Bigr]  e^{-i {\bf k} \cdot {\bf
x}},\nonumber\\
\pi^* (t, {\bf x}) &=& \int [d {\bf k}] \Bigl[ \dot{\varphi}^*_{\bf
k} (t) b^{\dagger}_{\bf k}(t) + \dot{\varphi}_{\bf k} (t) a_{\bf k}
(t) \Bigr]  e^{i {\bf k}. \cdot {\bf x}}.
\end{eqnarray}
Here $a_{\bf k}$ and $a^{\dagger}_{\bf k}$ are the annihilation and
creation operators for particles and $b_{\bf k}$ and
$b^{\dagger}_{\bf k}$ are those for antiparticles and $\varphi_{\bf
k}$ are auxiliary variables that will be determined later.

Then the Hamiltonian takes the form
\begin{eqnarray}
H (t) = \int [d {\bf k}]  \Bigl[ (\dot{\varphi}_{\bf k}^*
\dot{\varphi}_{\bf k} + \omega_{\bf k}^2 \varphi_{\bf k}^*
\varphi_{\bf k} ) (a_{\bf k}^{\dagger} a_{\bf k} + b_{\bf
k}^{\dagger} b_{\bf
k} + 1) \nonumber\\
+ (\dot{\varphi}_{\bf k}^{*2} + \omega_{\bf k}^2 \varphi_{\bf
k}^{*2} ) a_{\bf k}^{\dagger} b_{\bf k}^{\dagger} +
(\dot{\varphi}_{\bf k}^{2} + \omega_{\bf k}^2 \varphi_{\bf k}^{2} )
a_{\bf k} b_{\bf k} \Bigr], \label{ham os}
\end{eqnarray}
where
\begin{equation}
\omega^2_{\bf k} (t)  = (k_z + q A_z(t))^2 + {\bf k}_{\perp}^2 +
m^2. \label{omega}
\end{equation}
According to the invariant method
\cite{Lewis-Riesenfeld,Kim-Lee,Kim-Lee07}, we require the
time-dependent operators
\begin{eqnarray}
a_{\bf k} (t) &=& i \bigl[ \varphi^*_{\bf k} (t) \pi^*_{\bf k} -
\dot{\varphi}^*_{\bf k} (t) \phi_{\bf k} \bigr], \nonumber\\
a^{\dagger}_{\bf k} (t) &=& - i \bigl[ \varphi_{\bf k} (t) \pi_{\bf
k} - \dot{\varphi}_{\bf k} (t) \phi^*_{\bf k} \bigr], \label{a op}
\end{eqnarray}
and
\begin{eqnarray}
b_{\bf k} (t) &=& i \bigl[ \varphi^*_{\bf k} (t) \pi_{\bf k} -
\dot{\varphi}^*_{\bf k} (t) \phi^*_{\bf k} \bigr], \nonumber\\
b^{\dagger}_{\bf k} (t) &=& - i \bigl[ \varphi_{\bf k} (t)
\pi^*_{\bf k} - \dot{\varphi}_{\bf k} (t) \phi_{\bf k} \bigr],
\label{b op}
\end{eqnarray}
to satisfy the Liouville-von Neumann equation
\begin{eqnarray}
i \frac{\partial a_{\bf k} (t)}{\partial t} + [ a_{\bf k} (t), H
(t)] = 0, \nonumber\\
i \frac{\partial b_{\bf k} (t)}{\partial t} + [ b_{\bf k} (t), H
(t)] = 0.
\end{eqnarray}
They are satisfied only when $\varphi_{\bf k}$ is a complex solution
to the classical mode equation
\begin{eqnarray}
\ddot{\varphi}_{\bf k} + \omega_{\bf k}^2 \varphi_{\bf k} = 0.
\end{eqnarray}
We can fix the Wronskian conditions
\begin{equation}
\dot{\varphi}^*_{\bf k} (t) \varphi_{\bf k} (t) - \varphi^*_{\bf k}
(t) \dot{\varphi}_{\bf k} (t) = i, \label{wron}
\end{equation}
so that the standard equal-time commutation relations hold
\begin{eqnarray}
\bigl[ a_{ {\bf k}' } (t), a^{\dagger}_{\bf k} (t) \bigr] &=& \delta
( {\bf k}' - {\bf k} ), \nonumber\\ \bigl[ b_{ {\bf k}' } (t),
b^{\dagger}_{\bf k} (t) \bigr] &=& \delta ({\bf k}' - {\bf k} ).
\end{eqnarray}

The eigenstates of $a_{\bf k}$ and $b_{\bf k}$, invariant operators,
are exact quantum states of the time-dependent Schr\"{o}dinger
equation up to time-dependent phase factors \cite{Lewis-Riesenfeld}.
The ground state for each mode
\begin{eqnarray}
a_{\bf k} (t) \vert 0_{\bf k}; t \rangle = b_{\bf k} (t) \vert
\bar{0}_{\bf k}; t \rangle = 0, \label{gr st}
\end{eqnarray}
is an eigenstate with zero eigenvalue. The multi-particle and
antiparticle states are given by
\begin{eqnarray}
\vert n_{{\bf k}_1} \cdots; \bar{n}_{{\bf k}_2} \cdots; t \rangle =
\frac{a^{\dagger n_1}_{{\bf k}_1} (t)}{\sqrt{n_1!}} \cdots
\frac{b^{\dagger \bar{n}_2}_{{\bf k}_2} (t)}{\sqrt{\bar{n}_2!}}
\cdots \vert 0; t \rangle. \label{num st}
\end{eqnarray}
It is shown  in Ref. \cite{Kim-Page01} that the time-dependent phase
factors indeed vanish for the number states of $N_a = a_{\bf
k}^{\dagger} a_{\bf k}$ and $N_b = b_{\bf k}^{\dagger} b_{\bf k}$.
Thus the product of the ground states
\begin{eqnarray}
\vert 0; t \rangle = \prod_{\bf k} \vert 0_{\bf k}; t \rangle \vert
\bar{0}_{\bf k}; t \rangle,
\end{eqnarray}
is the time-dependent vacuum of the field, an exact quantum state.
In other words, the time-dependent vacuum is defined as
\begin{eqnarray}
a_{\bf k} (t) \vert 0; t \rangle = b_{\bf k} (t) \vert 0; t \rangle
= 0. \label{vac}
\end{eqnarray}

\section{Effective Action at Zero Temperature}

The effective action is determined by the scattering amplitude
(\ref{sc am})  between the ingoing vacuum and the outgoing vacuum.
To find the outgoing vacuum we need the Bogoliubov transformations
between the annihilation operators of the ingoing vacuum and the
outgoing vacuum:
\begin{eqnarray}
a_{{\bf k}, {\rm in}} &=& \mu^*_{\bf k} a_{{\bf k}, {\rm out}} -
\nu^*_{\bf k} b^{\dagger}_{{\bf k}, {\rm out}}, \nonumber\\
b_{{\bf k}, {\rm in}} &=& \mu^*_{\bf k} b_{{\bf k}, {\rm out}} -
\nu^*_{\bf k} a^{\dagger}_{{\bf k}, {\rm out}}, \label{in-out}
\end{eqnarray}
where
\begin{eqnarray}
\mu_{\bf k} &=& i \Bigl(\varphi^*_{\bf k} (\infty)
\dot{\varphi}^{\rm in}_{\bf k}
- \dot{\varphi}^*_{\bf k} (\infty) \varphi^{\rm in}_{\bf k} \Bigr), \nonumber\\
\nu_{\bf k} &=& i \Bigl(\varphi^*_{\bf k} (\infty)
\dot{\varphi}^{{\rm in}*}_{\bf k} - \dot{\varphi}^*_{\bf k} (\infty)
\varphi^{{\rm in}*}_{\bf k} \Bigr).
\end{eqnarray}
These coefficients satisfy the relation
\begin{eqnarray}
|\mu_{\bf k}|^2 - |\nu_{\bf k}|^2 = 1.
\end{eqnarray}
The inverse Bogoliubov transformations are
\begin{eqnarray}
a_{{\bf k}, {\rm out}} &=& \mu_{\bf k} a_{{\bf k}, {\rm in}} +
\nu^*_{\bf k} b^{\dagger}_{{\bf k}, {\rm in}}, \nonumber\\
b_{{\bf k}, {\rm out}} &=& \mu_{\bf k} b_{{\bf k}, {\rm in}} +
\nu^*_{\bf k} a^{\dagger}_{{\bf k}, {\rm in}}. \label{out-in}
\end{eqnarray}

To express the outgoing vacuum in terms of the particle states of
the ingoing vacuum, we may write the Bogoliubov transformation
(\ref{out-in}) as
\begin{eqnarray}
a_{{\bf k}, {\rm out}} = U (r_{\bf k}, \vartheta_{\bf k};
\theta_{\bf k}) a_{{\bf k}, {\rm in}} U^{\dagger} (r_{\bf
k}, \vartheta_{\bf k}; \theta_{\bf k}), \nonumber\\
b_{{\bf k}, {\rm out}} = U (r_{\bf k}, \vartheta_{\bf k};
\theta_{\bf k}) b_{{\bf k}, {\rm in}} U^{\dagger} (r_{\bf k},
\vartheta_{\bf k}; \theta_{\bf k}). \label{sq op}
\end{eqnarray}
Here $U$ is the evolution operator for each mode
\begin{eqnarray}
U (r_{\bf k}, \vartheta_{\bf k}; \theta_{\bf k}) =  S(r_{\bf k},
\vartheta_{\bf k}) P (\theta_{\bf k}),
\end{eqnarray}
where the overall phase factor and the two-mode squeeze operator are
\cite{Caves-Schumaker1,Caves-Schumaker2}
\begin{eqnarray}
P (\theta_{\bf k}) &=&  \exp \Bigl[i \theta_{\bf k}
\Bigl(a^{\dagger}_{{\bf k}, {\rm in}}a_{{\bf k}, {\rm in}} +
b^{\dagger}_{{\bf k}, {\rm in}} b_{{\bf k}, {\rm in}} + 1 \Bigr) \Bigr]\nonumber\\
S(r_{\bf k}, \vartheta_{\bf k}) &=& \exp \Bigl[ r_{\bf k} \Bigl(
a_{{\bf k}, {\rm in}} b_{{\bf k}, {\rm in}} e^{- 2i \vartheta_{\bf
k}} \nonumber\\ && ~~~~~~~~~~- a^{\dagger}_{{\bf k}, {\rm in}}
b^{\dagger}_{{\bf k}, {\rm in}} e^{2i \vartheta_{\bf k}} \Bigr)
\Bigr],
\end{eqnarray}
where the squeeze parameter $r_{\bf k}$, the squeeze angle
$\vartheta_{\bf k}$, and the overall phase angle $\theta_{\bf k}$
are determined by
\begin{eqnarray}
\mu_{\bf k} &=& e^{- i \theta_{\bf k}} \cosh r_{\bf k}, \nonumber\\
\nu_{\bf k}^* &=& - e^{-i \theta_{\bf k}} (e^{2 i \vartheta_{\bf k}}
\sinh r_{\bf k} ). \label{sq pa}
\end{eqnarray}

Then the outgoing vacuum is the two-mode squeezed state of the
ingoing vacuum
\begin{eqnarray}
\vert 0, {\rm out} \rangle = \prod_{\bf k} U (r_{\bf k},
\vartheta_{\bf k}; \theta_{\bf k}) \vert 0, {\rm in} \rangle,
\end{eqnarray}
from which follows the scattering amplitude
\begin{eqnarray}
\langle 0, {\rm out} \vert 0, {\rm in} \rangle = \prod_{\bf k} e^{ i
\theta_{\bf k}} \langle 0, {\rm in} \vert S^{\dagger} (r_{\bf k},
\vartheta_{\bf k}) \vert 0, {\rm in} \rangle.
\end{eqnarray}
Further, the squeeze operator can be factored as
\cite{Caves-Schumaker2}
\begin{eqnarray}
S(r_{\bf k}, \vartheta_{\bf k}) &=& \exp \Bigl[\xi_{\bf k}
a^{\dagger}_{{\bf k}, {\rm in}} b^{\dagger}_{{\bf k},
{\rm in}} \Bigr]\nonumber\\
&& \times  \exp \Bigl[ \frac{\gamma_{\bf k}}{2} \Bigl(
a^{\dagger}_{{\bf k}, {\rm in}} a_{{\bf k}, {\rm in}} +
b^{\dagger}_{{\bf k}, {\rm
in}} b_{{\bf k}, {\rm in}}+1 \Bigr) \Bigr] \nonumber\\
&&\times  \exp \Bigl[ - \xi_{\bf k}^* a_{{\bf k}, {\rm in}} b_{{\bf
k}, {\rm in}} \Bigr],
\end{eqnarray}
where
\begin{eqnarray}
\xi_{\bf k} &=& - e^{2 i \vartheta_{\bf k}} \tanh r_{\bf k},
\nonumber\\
\gamma_{\bf k} &=& \ln (1 - |\xi_{\bf k}|^2 ) = - 2 \ln (\cosh
r_{\bf k} ).
\end{eqnarray}
Thus the scattering amplitude is given by
\begin{eqnarray}
\langle 0, {\rm out} \vert 0, {\rm in} \rangle = \prod_{\bf k}
\frac{1}{\mu^*_{\bf k}}.
\end{eqnarray}
Finally, we find the effective action per unit volume
\begin{eqnarray}
{\cal L}_{\rm eff} = i \frac{qE}{(2 \pi)^3} \int d{\bf k}_{\perp}^2
\ln (\mu^*_{\bf k}), \label{eff-E}
\end{eqnarray}
where $qE/(2 \pi)$ is the number of states along the $z$-direction
and $1/(2 \pi)^2$ is the number of states for each ${\bf
k}_{\perp}$. It also follows that the decaying amplitude of the
ingoing vacuum is
\begin{eqnarray}
|\langle 0, {\rm out} \vert 0, {\rm in} \rangle |^2 = e^{ - 2 ({\rm
Im} {\cal S}_{\rm eff})} = e^{ - 2 \sum \ln |\mu_{\bf k}|},
\end{eqnarray}
where the summation is over all possible states in momentum and
spacetime.

Now we apply the formalism for the effective action to a constant
electric field with $A_z = - E t$. In a constant electric field, the
Klein-Gordon equation takes the form
\begin{eqnarray}
\Bigl[\partial_t^2 + m^2 + {\bf k}_{\perp}^2 + (k_z - qE t)^2 \Bigr]
\phi_{\omega, {\bf k}} (t) = 0.
\end{eqnarray}
The solution with the appropriate ingoing flux at $t = - \infty$ is
\begin{eqnarray}
\phi_{\omega, {\bf k}} (t) = D_p(z),
\end{eqnarray}
where $D_p (z)$ is the parabolic cylinder function \cite{GR} and
\begin{eqnarray}
z &=& \sqrt{\frac{2}{qE}} e^{i \pi/4} (k_z - qE t), \nonumber\\
p &=& - i \frac{m^2 + {\bf k}_{\perp}^2}{2 (qE)} - \frac{1}{2}.
\end{eqnarray}
It has the asymptotic form $D_p (z) \approx e^{- z^2/4} z^p$ for
$|z| \gg 1$. In the other region at $t = \infty$, the solution takes
another form \cite{GR2}
\begin{eqnarray}
D_p (z) = e^{i p \pi} D_{p} (-z) + \frac{\sqrt{2 \pi}}{\Gamma(-p)}
e^{i (p+1)\pi/2} D_{- p -1} (- iz). \label{con E-sol}
\end{eqnarray}
From Eq. (\ref{con E-sol}) we find the Bogoliubov coefficients
\begin{eqnarray}
\mu_{\bf k} &=& \frac{\sqrt{2 \pi}}{\Gamma(-p)} e^{i (p+1)\pi/2},
\nonumber\\
\nu_{\bf k} &=& e^{ i p \pi}.
\end{eqnarray}
It follows that
\begin{eqnarray}
| \mu_{\bf k} |^2 = e^{\frac{m^2 + {\bf k}_{\perp}^2}{2 qE}} + 1.
\end{eqnarray}
The effective action per unit volume and per unit time is then
\begin{eqnarray}
{\cal L}_{\rm eff} = i \frac{qE}{(2 \pi)^3} \int d {\bf k}_{\perp}^2
\Bigl[ \ln \sqrt{2 \pi} - \ln \Gamma (-p^*)  - i
\frac{(p^*+1)\pi}{2} \Bigr]. \label{eff-E2}
\end{eqnarray}
From the gamma function \cite{GR3}
\begin{eqnarray}
\ln \Gamma (z) = \int_0^{\infty} \Bigl[\frac{e^{-zt}}{1 - e^{-t}} -
\frac{e^{-t}}{1 - e^{-t}} + (z-1) e^{-t} \Bigr] \frac{dt}{t},
\label{gamma}
\end{eqnarray}
after doing the contour integral over the contour of a
quarter-circle in the first quadrant, we obtain
\begin{eqnarray}
\int_0^{\infty} \frac{e^{-zt}}{1 - e^{-t}} \frac{dt}{t} &=&
\frac{1}{2i} {\cal P} \int_0^{\infty} \frac{e^{ -(m^2+ {\bf
k}_{\perp}^2)s}}{\sin (qES)} \frac{ds}{s} \nonumber\\&& + \sum_{n =
1}^{\infty} \frac{(-1)^{n}}{2n} e^{- \frac{m^2+ {\bf
k}_{\perp}^2}{qE} n \pi},
\end{eqnarray}
where ${\cal P}$ denotes the principal value. Thus, integrating over
the momentum, the effective action is
\begin{eqnarray}
{\cal L}_{\rm eff} (E) &=& - \frac{qE}{16 \pi^2} {\cal P}
\int_0^{\infty} \frac{e^{- m^2 s}}{\sin (qE s)}\frac{ds}{s^2} +
\cdots + \nonumber\\&& + i \frac{qE}{16 \pi^3} \int d{\bf
k}_{\perp}^2 \ln (1+ e^{-\pi \frac{m^2+ {\bf k}_{\perp}^2}{qE}}).
\end{eqnarray}
Removing all the possible singular terms when $E = 0$, we finally
obtain the effective action (\ref{eff-E0}). Note that the second and
third term in Eq. (\ref{gamma}) and other terms in Eq.
(\ref{eff-E2}) have to removed by a regularization procedure, which
may have something to do with substraction of the zero-energy and
renormalization of charge. Our method recovers the standard result,
which can be compared with Eq. (3.23) of Ref. \cite{Ambjorn-Hughes},
where the imaginary part of the effective action gives the correct
pair production rate while the real part does not recover the
standard form.

\section{Conclusion}

In this paper we advanced a Hamiltonian method to calculate the
effective action in scalar QED not only in magnetic fields but also
electric fields. The widely adopted heat-kernel method is simple and
powerful to find the effective action of charged bosons and fermions
in a constant electromagnetic field. However, it is hard to make use
of it for inhomogeneous electric fields, though calculable in
principle. In the new method introduced in this paper, the effective
action is expressed in terms of the Bogoliubov coefficients only. It
is shown how the method works for a constant electric field. Though
the new method is more complicated than the heat-kernel method, it
recovers correctly the effective action from the solution of the
Klein-Gordon equation. It is expected that the new method may work
out the effective action in inhomogeneous electric fields as well.
It would be interesting to compare the effective action in
inhomogeneous electric fields with the resolvent method
\cite{Dunne-Hall} and also the derivative expansion method
\cite{Gusynin-Shovkovy}. The effective action at finite temperature
in inhomogeneous electromagnetic fields is in progress.

\acknowledgments

The authors thank Yongsung Yoon for useful comments. The work of
S.~P.~K. was supported by the Korea Research Foundation Grant funded
by the Korean Government (MOEHRD) (KRF-2007-C00167) and the work of
H.~K.~L. was supported by the Korea Science and Engineering
Foundation (KOSEF) grant funded by the Korea government (MOST) (No.
R01-2006-000-10651-0).
\appendix


\begin{thebibliography}{99}

\bibitem{Schwinger} J.~Schwinger, Phys. Rev. {\bf 82}, 664 (1951).

\bibitem{Heisenberg-Euler} W.~Heisenberg and H.~Euler, Z. Physik {\bf 98}, 714
(1936).

\bibitem{Weisskopf} V.~Weisskopf, K. Dan. Vidensk. Selsk. Mat. Fys. Medd. {\bf
XIV}, 6 (1936).

\bibitem{Dunne} G.~V.~Dunne, ``Heisenberg-Euler Effective Lagrangians: Basics
and Extensions,'' {\it From Fields to Strings: Circumnavigating
Theoretical Physics}, edited by M.~Shifman, A. Vainshtein, and J.
Wheater, (World Scientific, Singapore, 2005), Vol. I, pp. 445-522,
hep-th/0406216.

\bibitem{Ringwald} A. Ringwald, Phys. Lett. B {\bf 510}, 107 (2001);
``Fundamental Physics at an X-Ray Free Electron Laser,'' {\it
Electromagnetic Probes of Fundamental Physics}, edited by
W.~Marciano and S.~White (World Scientific, Singapore, 2003), pp.
63-74, hep-ph/0112254; ``Boiling the Vacuum with an X-Ray Free
Electron Laser,'' {\it Quantum Aspects of Beam Physics 2003}, edited
by P.~Chen and K.~Reil (World Scientific, Singapore, 2004), pp.
149-163, hep-ph/0304139.

\bibitem{Wood} P.~M.~Wood and C.~Thompson, ``Soft Gamma Repeaters
and Anomalous X-ray Pulsars: Magnetar Candiates,'' in {\it Compact
Stellar X-ray Sources}, edited by W.~H.~G.~Lewin and M.~van~der~Klis
(Cambridge Univ., Cambridge, 2006), pp 547-586, astrop-ph/0406133.

\bibitem{Duncan} R.~C.~Duncan, ``Physics in Ultra-strong Magnetic
Fields,'' astro-ph/0002442 (2000).

\bibitem{Harding-Lai} A.~K.~Harding and D.~Lai, Rept. Prog. Phys.
{\bf 69}, 2631 (2006).

\bibitem{Ito} N.~Itoh, Prog. Theor. Phys. {\bf 44}, 291 (1071).

\bibitem{Bodmer} A.~R.~Bodmer, Phys.~Rev.~D {\bf 4}, 1601 (1971).

\bibitem{Witten} E.~Witten, Phys. Rev. D {\bf
30}, 272 (1984).

\bibitem{Farhi} E.~Farhi and R.~L.~Jaffe, Phys. Rev. D {\bf 30}, 2379 (1984).

\bibitem{Alcock} C.~Alcock, E.~Farhi, and A.~V.~Olinto, Astrophys.
J. {\bf 310}, 261 (1986).

\bibitem{Haensel} P.~Haensel, J.~L.~Zdunik, and R.~Schaeffer,
Astron. Astrophys. {\bf 160}, 121 (1986).

\bibitem{Usov98} V.~V.~Usov, Phys. Rev. Lett. {\bf 80}, 230 (1998).

\bibitem{Usov05} V.~V.~Usov, T.~Harko, and K.~S.~Cheng, Astrophys.
J. {\bf 620}, 915 (2005).

\bibitem{Harko06} T.~Harko and K.~S.~Cheng, Astrophys. J. {\bf 643},
318 (2006).

\bibitem{Kim-Lee07} S.~P.~Kim and H.~K.~Lee, Phys. Rev D {\bf 76},
125002 (2007).

\bibitem{Lewis-Riesenfeld} H.~R.~Lewis,~Jr. and W.~B. Riesenfeld, J.
Math. Phys. {\bf 10}, 1458 (1969).

\bibitem{Ambjorn-Hughes} J.~Ambj{o}rn, R.~J.~Hughes, and
N.~K.~Nielsen, Ann. Phys. {\bf 150}, 92 (1983).

\bibitem{wybourne} B.~G.~Wybourne, {\it Classical Groups for
Physicists} (A Wiley-Interscience Pub., New York, 1974), pp.
268-275.

\bibitem{greiner94} W.~Greiner and J.~Reinhardt, {\it Quantum
electrodynamics} 2nd ed. (Springer, Berlin, 1994), pp. 376-378.

\bibitem{FHM} K.~Freese, C.~T.~Hill, and M.~Mueller, Nucl. Phys. B
{\bf 255}, 693 (1995).

\bibitem{Kim-Lee} S.~P.~Kim and C.~H.~Lee, Phys. Rev. D {\bf 62},
125020 (2000).

\bibitem{Kim-Page01} S.~P.~Kim and D.~N.~Page, Phys. Rev. A {\bf 64},
012104 (2001).

\bibitem{Caves-Schumaker1} C.~M.~Caves and B.~L.~Schumaker, Phys.
Rev. A {\bf 31}, 3068 (1985).

\bibitem{Caves-Schumaker2} B.~L.~Schumaker and C.~M.~Caves, Phys. Rev. A {\bf 31}, 3093
(1985).

\bibitem{GR} I.~S.~Gradshteyn and I.~M.~Ryzhik, {\it Table of
Integrals, Series, and Products} (Academic Press, San Diego, 1994).

\bibitem{GR2} I.~S.~Gradshteyn and I.~M.~Ryzhik, {\it Table of
Integrals, Series, and Products} (Academic Press, San Diego, 1994),
9.248-3.

\bibitem{GR3} I.~S.~Gradshteyn and I.~M.~Ryzhik, {\it Table of
Integrals, Series, and Products} (Academic Press, San Diego, 1994),
8.341-3.

\bibitem{Dunne-Hall} G.~Dunne and T.~Hall, Phys. Rev. D {\bf 58},
105022 (1998).

\bibitem{Gusynin-Shovkovy} V.~P.~Gusynin and I.~A.~Shovkovy, J.
Math. Phys. {\bf 40}, 5406 (1999).

\end{thebibliography}
\end{document}